\documentclass[a4paper,11pt,DIV=calc]{article}

\usepackage[a4paper,left=30mm,right=30mm,top=30mm,bottom=30mm,marginpar=23mm]{geometry}

\usepackage[backref=page]{hyperref}
\renewcommand*{\backref}[1]{}
\renewcommand*{\backrefalt}[4]{%
    \ifcase #1 (Not cited.)%
    \or        (Cited on page~#2.)%
    \else      (Cited on pages~#2.)%
    \fi}
\usepackage{subfig}
\usepackage{amsmath,amssymb}
\usepackage{bbm}
\usepackage{amsthm}
\usepackage{pgfplots}
\usepackage{graphicx}
\usepackage{MnSymbol}
\usepackage{color}
\newcommand{\no}[3]{\ensuremath{(#1\,#2\,#3)_\gamma}}
\newcommand{\ve}[3]{\ensuremath{[#1\,#2\,#3]_\gamma}}

\newcommand{\noa}[3]{\ensuremath{(#1\,#2\,#3)_\alpha}}
\newcommand{\vea}[3]{\ensuremath{[#1\,#2\,#3]_\alpha}}

\newcommand{\noc}[3]{\ensuremath{(#1,#2,#3)}}
\renewcommand{\vec}[3]{\ensuremath{[#1,#2,#3]}}

\newcommand{\vecomma}[3]{\ensuremath{(#1,#2,#3)_\gamma}}

\theoremstyle{plain} \newtheorem{Theorem}{Theorem}
\theoremstyle{plain} \newtheorem{Proposition}[Theorem]{Proposition}
\theoremstyle{definition} 
\theoremstyle{plain} \newtheorem{Lemma}{Lemma}

\newcommand{\id}{{\mathbb{I}}}

\newcommand{\la}{\lambda}

\newcommand{\tp}{\otimes}

\newcommand{\bb}{\ensuremath{\mathbf{b}}}
\newcommand{\cc}{\ensuremath{\mathbf{c}}}
\newcommand{\mm}{\ensuremath{\mathbf{m}}}
\newcommand{\kk}{\ensuremath{\mathbf{k}}}

\newcommand{\al}{\alpha}

\newcommand{\ga}{\gamma}

\mathchardef\mdash="2D

\newcommand{\PP}{\mathcal{P}^{24}}

\newcommand{\aaa}{\mathbf{a}}

\newcommand{\ee}{\mathbf{e}}
\newcommand{\nnn}{\mathbf{n}}
\newcommand{\nn}{\mathbf{n}}
\newcommand{\dd}{\mathbf{d}}
\newcommand{\pp}{\mathbf{p}}

\newcommand{\aand}{\mbox{ and }}
\newcommand{\PT}{\ensuremath{\mathrm{P}}}

\newcommand{\GT}{\ensuremath{\mathrm{GT}}}
\newcommand{\KS}{\ensuremath{\mathrm{KS}}}
\newcommand{\NW}{\ensuremath{\mathrm{NW}}}

\renewcommand{\vec}[3]{\begin{bmatrix}#1 \\ #2\\ #3 \end{bmatrix}}
\renewcommand{\noc}[3]{\begin{pmatrix}#1 \\ #2\\ #3 \end{pmatrix}}

\newcommand{\fcc}{{\ensuremath{\mathrm{fcc}}}}
\newcommand{\bcc}{{\ensuremath{\mathrm{bcc}}}}
\newcommand{\mT}{^{\mathrm{-T}}\mkern-1mu}
\newcommand{\T}{^{\mathrm{T}}\mkern-2mu}

\newcommand{\one}{{\mathrm{I}}}
\newcommand{\two}{{\mathrm{II}}}

\usepackage{cleveref}
\usepackage{authblk}
\usepackage{overpic}

\usepackage{fancyhdr}
\newcommand{\nono}[3]{\ensuremath{\lbrace#1\,#2\,#3\rbrace_\gamma}}

\begin{document}

\makeatletter
\def\@maketitle{%
  \newpage
  \null
  \vskip 2em%
  \begin{center}%
  \let \footnote \thanks
    {\Large\bf  \@title \par}%
    \vskip 1.5em%
    {\normalsize
      \lineskip .5em%
      \begin{tabular}[t]{c}%
        \@author
      \end{tabular}\par}%
    \vskip 1em%
    {\normalsize \@date}%
  \end{center}%
  \par
  \vskip 1.5em}
\makeatother
\title{{A Parameter Free Double Shear Theory for Lath Martensite}}

\author{Konstantinos Koumatos%
  \thanks{\texttt{k.koumatos@sussex.ac.uk}}}
\affil{\small\textit{Department of Mathematics,}\\ \textit{University of Sussex,} \\ \textit{Pevensey 2 Building,} \\ \textit{Brighton BN1 9QH, UK}}

\author{Anton Muehlemann%
  \thanks{\texttt{amuehle@berkeley.edu}}}
\affil{\small\textit{Department of Civil Engineering,}\\ \textit{University of California Berkeley,}\\\textit{Davis Hall,} \\ \textit{Berkeley CA 94720, USA}}

\date{Dated: \today}

\maketitle

 \hypersetup{
  pdfauthor = {Konstantinos Koumatos (University of Sussex) and Anton Muehlemann (University of California Berkeley)},
  pdftitle = {A Parameter Free Double Shear Theory for Lath Martensite},
  pdfsubject = {MSC (2010): 74N05, 74N15},
  pdfkeywords = {lath martensite, microstructure, twins within twins, (557) habit planes, double shear theories, energy minimization, non-classical interfaces}
}

\vspace{4pt}
\begin{abstract}
A double shear theory is introduced that predicts the commonly observed \nono{5}{5}{7} habit planes in low-carbon steels. The novelty of this theory is that {no parameter fitting is necessary. Instead,} the shearing systems are chosen in analogy to the original (single shear) phenomenological theory of martensite crystallography as those that are macroscopically equivalent to twinning. Out of all the resulting double shear theories, the ones leading to certain \nono{h}{h}{k} habit planes naturally arise as those having small shape strain magnitude and satisfying a condition of maximal compatibility, thus making any parameter fitting unnecessary. An interesting finding is that the precise coordinates of the predicted \nono{h}{h}{k} habit planes depend sensitively on the lattice parameters of the fcc (face-centered cubic) and bcc (body-centered cubic) phases. Nonetheless, for various realistic lattice parameters in low carbon steels, the predicted habit planes are near \nono{5}{5}{7}. The examples of Fe-$0.252$C and Fe-$0.6$C are analyzed in detail along with the resulting orientation relationships which are consistently close to the Kurdjumov-Sachs model. Furthermore, a MATLAB app \href{https://github.com/AntonMu/LathApp}{\emph{``Lath Martensite''}} is provided which allows the application of this model to any other material undergoing an fcc to bcc transformation.  
\end{abstract}

\noindent\textsc{MSC (2010): 74A05, 74N05, 74N10} 

\noindent\textsc{Keywords:}  Nishiyama-Wassermann, Kurdjumov-Sachs, tetragonal, orientation relationships, transformation strain, steel, fcc to bcc, fcc to bct, Pitsch, Greninger-Troiano, Bain, Inverse Greninger-Troiano

\tableofcontents

\section{Introduction}
\begin{figure}[ht]
\centering
\includegraphics[width=.9\textwidth]{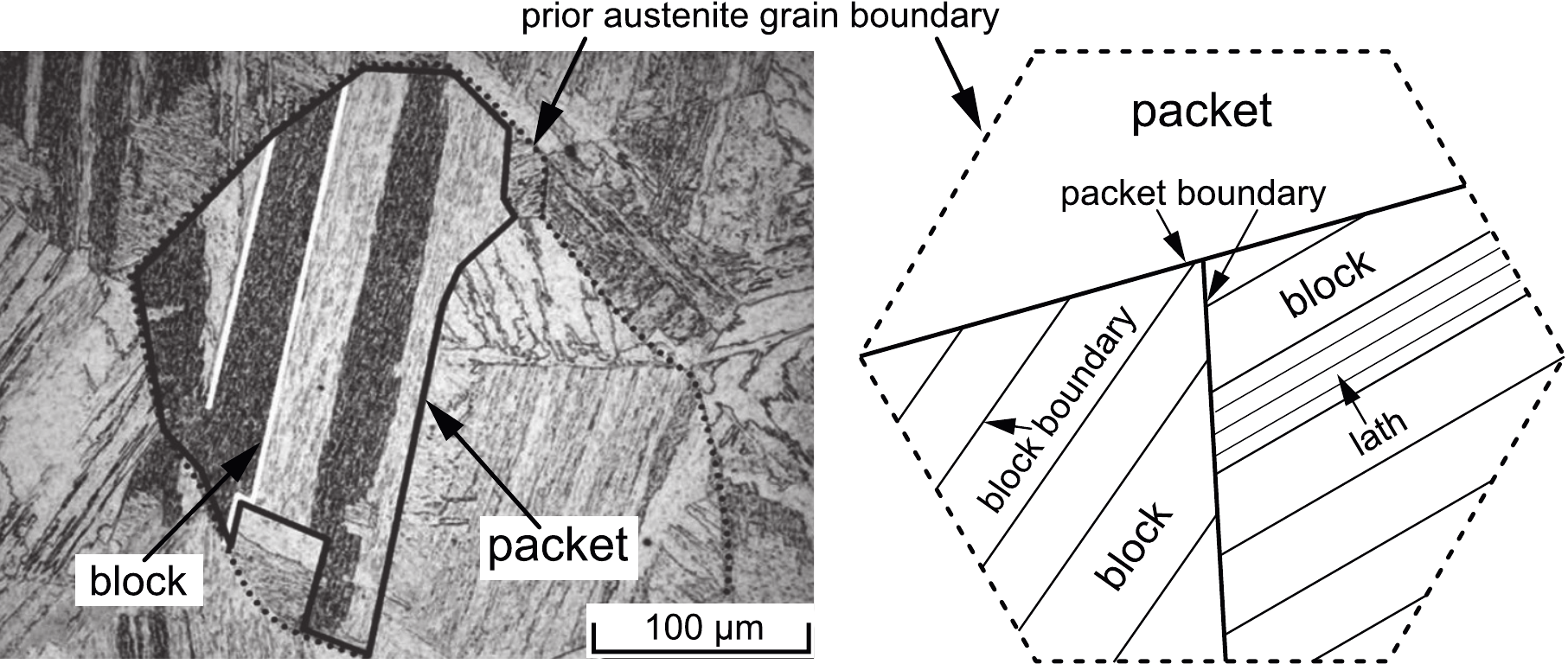}
\caption{A micrograph and schematic representation of lath martensite within a prior austenite grain (after \cite{Morito}). Each packet is divided into layers of blocks, each consisting of layers of laths.}\label{fig:lath}
\end{figure}

Among the various morphologies of martensite in ferrous materials, lath martensite is one of the most important and well studied owing to its significant industrial applications. Studies on the morphology of lath martensite have revealed that it possesses a hierarchical structure. The prior austenite grain is divided into packets each containing blocks of laths sharing the same orientation, see Figure \ref{fig:lath}.

The length scale of individual laths is typically very fine and cannot be observed under the optical microscope. Nevertheless, many properties of steel, such as its strength and toughness, depend on the effective grain size, see \cite{Morito}, and a good understanding of its morphology is hence crucial.

The dominant theoretical models for the description of lath morphology are double shear theories, e.g. \cite{Ross,Kelly}, according to which the overall shape strain $F$ can be decomposed as
\begin{equation}\nonumber
 F=RBST,
\end{equation}
where $R$ is a rigid body rotation, $B$ is one of the Bain strains and $S,T$ are two shearing systems (see Section \ref{Secmodel} for details). In addition, $F$ is required to be an invariant plane strain, henceforth abbreviated to IPS, i.e. $F$ is required to be of the form $F=\id+\cc \tp \pp$ leaving a plane with normal $\pp$ undistorted. In the literature (\cite{Kelly,Ross,Morito,Kinney,Maresca}) many different choices of shearing systems $S$ and $T$ have been proposed to explain the morphology of lath martensite and have been successful in explaining many of the macroscopic observables, such as the typically observed $\nono{5}{5}{7}$, $\nono{2}{2}{3}$ or $\nono{1}{1}{1}$ habit planes, or the observed orientation relationships typically between Kurdjumov-Sachs (\KS) and Nishiyama-Wasserman (\NW). A common feature of phenomenological theories in lath martensite \cite{Kelly,Ross} is that a priori many different choices of shearing systems need to be considered. This choice is usually driven by physical intuition or experimental input. Hence, the identification of the correct systems requires parameter fitting and/or parameter estimation to comply with observables. This approach is in sharp contrast to the (original) phenomenological theory of martensite crystallography (PTMC) \cite{Read,BM123} where a \emph{single} shear theory was proposed, and was very successful in predicting the morphology of plate martensite. In this single shear theory, the system was chosen as the \emph{unique} shear arising from twinning in martensite.

More recently, Maresca \& Curtin \cite{Maresca} have proposed a double shear model in which the choice of shearing systems is driven by robust atomistic simulations incorporating the experimentally observed stepped interface with step direction along $[-1, 0, 1]_{\gamma}$. In their theory, any orientation relationship can be incorporated in the model and their free parameter - one of the shearing magnitudes - is fixed by the IPS requirement, providing good agreement with observed habit planes.

However, a common feature of many of these theories is that a priori, based on physical intuition, different choices of shearing systems need to be considered.\footnote{{Among the exceptions lies the recent work of \cite{Maresca} where robust atomistic simulations drove the choice of one of the two shearing systems.}} As a result, the identification of the correct systems {requires parameter fitting and/or parameter estimation.} This approach is in sharp contrast to the (original) phenomenological theory of martensite crystallography (PTMC) \cite{Read,BM123} where a \emph{single} shear theory was proposed, and was very successful in predicting the morphology of plate martensite. In this single shear theory, the system was chosen as the \emph{unique} shear arising from twinning in martensite.
 
Other notable models for the prediction of lath morphology include the work of Baur, Cayron, Log\'e \cite{Cayron} as well as the careful search among many different morphologies performed by Qi, Khachaturyan \& Morris \cite{QKM}. In \cite{Cayron}, the authors reach near $\nono{5}{5}{7}$ habit planes by averaging two \KS variants\footnote{The computation of the \KS and \NW variants as rotated Bain strains is performed in the spirit of the work by Jawson \& Wheeler \cite{Jawson48}, see also \cite{KSPaper}.} sharing the same Bain axis to produce an \NW variant whose cofactor matrix has an eigenvector $(1,\sqrt{2},1)_\gamma$, i.e. it leaves the $(1,\sqrt{2},1)_\gamma$ plane invariant.

In a different spirit, Qi, Khachaturyan \& Morris \cite{QKM}, explain the lath morphology by investigating various possible transformations of increasing complexity: a single-variant plate with a single slip system, a single variant plate with two slip systems (akin to double shear theories), a composite plate of two variants each with a single slip system, as well as a composite plate of two variants each with two slip systems. We note that the latter two morphologies resemble twinning between slipped regions in martensite. A wide search through thousands of possible shears leads the authors to construct double shear theories with near $\nono{5}{5}{7}$ habit planes. Similarly, under the assumption of double slip and composite plates, they construct theories with near $\nono{1}{1}{1}$ habit planes involving two variants that share a common Bain axis with a \KS orientation relationship to austenite (cf. \cite{Cayron}).

{In the present article, we extend upon previous results in \cite{NewPersp} and build double shear theories to describe the deformation in a single lath by choosing both shearing systems as those which are macroscopically equivalent to twinning. Based on this assumption, we are able to build a model that predicts double shear theories with certain $\nono{h}{h}{k}$ habit planes as those satisfying a condition of low shape strain magnitude and maximal compatibility.} {We extend our previous work in several important ways: \begin{itemize}
\item we investigate the resulting orientation relationships and compare them to existing experimental work,
\item we include a detailed discussion on determining appropriate lattice parameters for the transformation by taking into account thermal expansion,
\item we determine the effect of volume change on, among other things, the observed habit plane normals and orientation relationships, 
\item we provide a \href{https://github.com/AntonMu/LathApp}{MATLAB app} with a user friendly interface which allows anyone to easily calculate all quantities of interest based on the lattice parameters, and 
\item 
we follow a different philosophy which is closer to the existing literature on existing double shear theories.
\end{itemize}
}

We remark that some of the calculations on twinning performed in the present article resemble those typical of elasticity based models of phase transformations, e.g. plate-martensite \cite{Read,BM123} or shape-memory alloys \cite{Bha,BallFine}. However, we do not assume any such elasticity model.

Though twinning in martensite is used to derive the shear deformations in our double shear theories, our deformations are only \emph{macroscopically} equivalent to twinning and do not necessitate, or suggest, that twinning is the internal structure. Indeed, the internal structure admits various interpretations such as double twinning in martensite, a twinned region between two variants each with an active slip system, or a single-variant with two active slip systems, the slip systems being of an unconventional nature. Such alternative interpretations resemble those of \cite{QKM} and we refer the reader to Sections \ref{section:singleshear}, \ref{section:doubleshear} for further elaboration.

The article is structured as follows: in Section \ref{Secmodel}, we introduce our model for lath martensite. As the phase transformation from austenite to martensite occurs within a large temperature interval, thermal expansion effects become significant and need to be taken into account. To this end, we discuss material parameters for our model based on previous experimental estimations of thermal expansion coefficients for various ferrous materials. We then briefly review single/double shear theories and introduce our method of selecting shearing systems that are macroscopically compatible with (double) twinning. This selection mechanism results in two one-parameter families of admissible habit planes $\pp$ with the property that, for each normal, there is a shape strain $F=RBST$ leaving $\pp$ invariant, i.e.
\[
F=\id+\cc \tp \pp.
\]
We conclude Section \ref{Secmodel} by introducing a selection mechanism for the produced shape strains based on low shape strain magnitude and a criterion of \emph{maximal compatibility}. Surprisingly, under the above criteria and for appropriate lattice parameters, the fcc (face-centered cubic) to bcc (body-centered
cubic) transformation results in habit planes very close to $\nono{5}{5}{7}$.

In Section \ref{Secresults}, we apply our model to Fe-$0.252$C and Fe-$0.6$C and give the precise form of the resulting shearing systems and invariant plane strains. In both cases, the resulting habit planes are near $\nono{5}{5}{7}$. For these double shear theories, we also compute the resulting orientation relationships and show that they are close to a variant of the Kurdjumov-Sachs orientation relationship. A similar analysis for any other choice of material parameters can be performed with the MATLAB App \emph{``Lath Martensite''} available at\texttt{\href{https://github.com/AntonMu/LathApp}{github.com/AntonMu/LathApp}}.

\section{Model}\label{Secmodel}
The transformation from fcc $\gamma$-austenite to bcc $\alpha$-martensite can be described by the three different Bain strains, see e.g. \cite{Bain,OptLat}, given by  
\begin{equation}\label{BainStrains}
B_1 = \frac{a_\bcc}{a_\fcc} \begin{pmatrix} 1  & 0 & 0 \\ 0 & \sqrt{2} & 0\\ 0 & 0 & \sqrt{2} \end{pmatrix}, 
B_2 = \frac{a_\bcc}{a_\fcc}\begin{pmatrix} \sqrt{2}  & 0 & 0 \\ 0 & 1 & 0\\ 0 & 0 & \sqrt{2} \end{pmatrix}, 
B_3 = \frac{a_\bcc}{a_\fcc} \begin{pmatrix} \sqrt{2}  & 0 & 0 \\ 0 & \sqrt{2} & 0\\ 0 & 0 & 1 \end{pmatrix},
\end{equation}
where $a_\fcc$ and $a_\bcc$ denote the lattice parameters of the $\gamma$ and $\alpha$ phases, respectively. We note that the $B_i$'s are simply the stretch components of the transformation in the $\ga$-basis and may be followed by an arbitrary rigid body rotation $R$. That is, any strain of the form $RB_i$ still transforms fcc austenite to bcc martensite. Depending on the specific rotation, the corresponding orientation relationship could be Kurdjumov-Sachs, Nishiyama-Wassermann or others \cite{KSPaper}.

\subsection{Thermal Expansion}\label{SecThermal}
In our proposed double shear theory, the only variable input parameters are $\la$ (equivalent to the shearing magnitude of one of the shears, see \eqref{IPS}) and the lattice parameters of the $\gamma$ and $\alpha$ phases which determine the precise form of the Bain strains $B_i$. It turns out that the habit planes (see the MATLAB App \href{https://github.com/AntonMu/LathApp}{\emph{``Lath Martensite''}}) depend sensitively on these lattice parameters.
As is evident from \eqref{BainStrains} it is in fact only the ratio, ${a_\bcc}/{a_\fcc}$, of the fcc and bcc lattice parameters that determines the Bain strains which can be translated into a single volume change parameter $\Delta V=\det B_i-1=2\, ({a_\bcc}/{a_\fcc})^3-1$.

Since the austenite to martensite transformation spans a large temperature interval, thermal expansion can have a significant effect on the parameter $\Delta V$. To take thermal expansion into account, we ought to consider $a_\bcc$ and $a_\fcc$ as functions of temperature $T$. Instead of specifying a single transformation temperature, here, we take into account all temperatures where both phases coexist, i.e. the temperature interval between the martensite start $M_s$ and finish $M_f$ temperature. The volume change parameters $\Delta V(T)=({a_\bcc(T)}/{a_\fcc(T)})^3-1$ are then considered for fcc and bcc lattice parameters at a common temperature $T$ within that interval. In our opinion such an approach seems more suitable than taking $a_\fcc$ at $M_s$ and $a_\bcc$ at $M_f$ which seems common in the literature.

It has been difficult to find reliable data on thermal expansion that includes both the austenite and martensite phase for the same material and spans a wide enough temperature range. For slowly cooled carbon steels, thermal expansion data for $\ga$ austenite and $\al$ ferrite can for instance be found in \cite{Onink} and \cite{FewSteels}. The transformation temperatures for these materials are around $1000$K and thus different from typical transformation temperatures reported for austenite to martensite transformations which, according to \cite{Steels}[Fig. 5.18], are between $400$K and $800$K. However, due to the lack of accurate data for martensite and the fact that both martensite and ferrite are bcc, we base our calculations on \cite{Onink, FewSteels}. {This approach is similar to the strategy used in \cite{Morito}. }

By \cite{Onink}, for Fe-0.6C, the austenite lattice parameters in the interval $1030$ K to $1250$ K are given by $a_\fcc(T)=0.36511\, (1 + 23.3 \cdot 10^{-6} (T-1000))$ and the ferrite lattice parameters in the interval $800$ K to approximately $1050$ K by $a_\bcc(T) = 0.28863 (1 + 17.5 \cdot 10^{-6} (T-800))$. Thus, the volume change in the temperature range of coexistence is approximately $-0.2\%$. Similarly, using the corresponding formulas for Fe-0.4C we obtain a volume change of approximately $0.35\%$ at $1050$ K. 

In \cite{FewSteels}, the sample S556 with composition Fe-0.252C\footnote{Alloying elements 0.06Mn, 0.012P, 0.035S, 0.007Si.} transforms from fcc to bcc between $1079$ K and $1005$ K. Using graphical extrapolation from the corresponding graph of thermal expansion, \cite{FewSteels}[Fig. 8], gives a volume change between $0.76\%$ and $0.97\%$. Similarly for the sample Fe-0.35C\footnote{Alloying elements 1.42Mn, 0.013P, 0.057S, 0.20Si, 1.00Cr, 0.11Vr.} we obtain a volume change between $0.6\%$ and $0.64\%$ for temperatures between $955$K and $934$K. 

\subsection{Single shear theories}
\label{section:singleshear}
An important and natural assumption in the phenomenological theory of martensite crystallography (PTMC) \cite{Read,BM123}, is that the shape strain $F$ is an invariant plane strain (IPS), i.e. $F=\id + \cc \tp \pp$ leaves a plane of normal $\pp$ invariant. It is well known that a simple Bain deformation of the form $F=RB_i$ cannot be an IPS. However, if one allows simple or multiple shears of the original lattice, the total shape strain can become an IPS.

In the seminal work of \cite{Read,BM123} a simple shear theory was proposed to accurately predict the features of plate martensite. An element that made their theory so successful was a unique way of choosing their shearing system. Instead of allowing all lattice invariant shearing systems they only allowed the ones that arise from twinning in martensite. We note that, on a macroscopic scale, one cannot distinguish between internal twinning and other structures resulting in the same macroscopic strain, such as slip. 

Below we recall Mallard's law as a convenient way of finding twinning systems.

\begin{Lemma}{(Mallard's law)}\label{LemmaMal}\\
 Let $A$ be a $3\times3$ matrix, $P=-\id + 2 \ee \tp \ee$ be a $180^\circ$ rotation about the unit vector $\ee$ and $B=PAP$. Then the equation $RB = A + \aaa \tp \nnn$ admits two solutions $(R^\one,\aaa^\one,\nnn^\one)$ and $(R^\two,\aaa^\two,\nnn^\two)$ given by
\begin{align} \tag{I}\label{EqM1}
  \aaa^\one&=2 \left(\frac{A\mT \ee}{|A\mT \ee|^2}-A\ee\right), \ &&\nnn^\one=\ee,\\ 
   \aaa^\two&=\frac{2N}{|A\ee|^2}A\ee, \ &&\nnn^\two=\frac{1}{N}\left(|A\ee|^2 \ee -A\T A\ee\right), \tag{II} \label{EqM2}
\end{align}
where $N$ is chosen such that $\nnn^\two$ is of unit length. The unknown rotations $R^\one$ and $R^\two$ can by calculated as $R= (A+ \aaa \tp \nnn)B^{-1}$. In particular, the strains $RB$ and $A$ are twin related.
\end{Lemma}

In our case, the axes of the necessary $180^\circ$ rotations $P$ relating the Bain variants, i.e. $B_i=PB_jP$, are all in the family \nono{1}{1}{0} and by Mallard's law, for each pair of Bain variants $B_i$ and $B_j$, there are two solution triples $(R^\one,\aaa^\one,\nnn^\one)$ and $(R^\two,\aaa^\two,\nnn^\two)$ such that 
\begin{equation}\label{EqTwinning}
R^{\one}B_j-B_i=\aaa^{\one} \tp \nn^{\one} \aand R^{\two}B_j-B_i=\aaa^{\two} \tp \nn^{\two}.
\end{equation}
In addition, all the resulting twins are of compound type, i.e. there exist two different $180^\circ$ rotations (with axes $\ee_1$ and $\ee_2$) relating $B_i$ and $B_j$. In particular, the first solution \eqref{EqM1} of Mallard's law for $P=-\id + 2 \ee_1 \tp \ee_1$ coincides with the second solution \eqref{EqM2} of Mallard's law for $P=-\id + 2 \ee_2 \tp \ee_2$. Put differently, by considering all $180^\circ$ rotations relating $B_i$ and $B_j$, it suffices to only consider the first solution of Mallard's law.\footnote{The family of all possible axes is precisely \nono{1}{1}{0}. } Therefore, we can, without loss of generality, only consider the first solution \eqref{EqM1} in Mallard's law and suppress the superscript.

Condition \eqref{EqTwinning} implies that the two strains $B_i$ and $RB_j$ can be internally twinned across an interface with normal $\nnn$ resulting in a macroscopic strain F of the form $F=(1-\la) B_i+\la RB_j$, where $\la$ is the volume fraction of $RB_j$ in the twinning system. Using \eqref{EqTwinning}, the macroscopic strain $F$ can be expressed as
\begin{alignat}{3}\label{EqSingle}
 F=B_i+\la (RB_j-B_i)  = B_i+\la \aaa \tp \nn = B_i(\id+\la B_i^{-1}\aaa \tp \nn)=:B_iS_{ij}(\la),
\end{alignat}
where $S_{ij}(\la)=\id+\la B_i^{-1}\aaa \tp \nn$ is a shear with shearing magnitude $g=|\la B_i^{-1}\aaa|$, shearing direction $\dd=\la B_i^{-1}\aaa/g$ and shearing normal $\nnn$. In particular, two internally distinct states, twinning in the product lattice and slip in the parent followed by the phase transformation, can result in the same macroscopic strain $F$.

In the (single shear) PTMC this observation was crucial in determining the possible shearing system. Following \cite{Read,BM123}, 
the parameter $\lambda$ is then fully determined by the requirement that the total shape strain $RF$, for some rotation $R$, is an IPS. 

We remark that the condition $RB = A + \aaa \tp \nnn$ - that is, $RB$ and $A$ are \emph{rank-one connected} - is \emph{necessary} for a deformation taking the values $RB$ and $A$ on either side of an interface with normal $\nn$ to remain continuous. Thus, it is a requirement pertaining not only to twinning but any deformation where there is no breaking of atomic bonds. Indeed, typical slip systems can be derived as lattice-invariant shearing deformations that are rank-one connected to the undistorted lattice. We refer the reader to \cite{Inamura13} or \cite{MiyamotoKC} where attainment of the rank-one condition is used as a criterion of kinematic compatibility.

\subsection{Double shear theories}\label{section:doubleshear}

The (original) PTMC has proven very successful in explaining features of various martensitic transformations. However, when applied to steels it was only successful in a few cases, such as the $\no{3}{10}{15}$ transformation (see \cite{Wata1,Efsic,Dunne}), and failed to give adequate predictions for e.g. the $\no{2}{2}{5}$ (see  \cite{Morton,WaymanCr,Dunne}) or the $\no{5}{5}{7}$ transformations. A good overview of the original theory and its various extensions can e.g. be found in \cite{DunneWayman}.

In the case of lath martensite, widely used extensions of the PTMC are double shear theories, e.g. \cite{Kelly,Ross}. In a double shear theory the total shape strain $F$ can be expressed as
\begin{equation}\label{FDouble}
 F = R B_i S T,
\end{equation}
for two shearing systems of the form $S=\id + \aaa \tp \nnn$ and $T=\id + \bb\tp \mm$. As in the single shear theory, the goal is to find shearing systems $S$ and $T$ and a rotation $R$ such that the total shape strain $F$ becomes an invariant plane strain (IPS). We observe that if one regards $B_iS$ as a variant itself, then $F$ from \eqref{FDouble} is related to this new variant by the single shear $T$. Since, as seen in the calculation leading to \eqref{EqSingle}, there is a one-to-one connection between single shears and simple twins, there is a similar connection between shears of shears (double shears) and twins of twins (double twins). 

Owing to the symmetry of the Bain variants, it is easy to show that for any $\la\in[0,1]$, the six possible shape strains arising from simple twins (or shears) are again related by $180^\circ$ rotations about vectors in the family \nono{1}{1}{0}. Hence, we can once again apply Mallard's law to the sheared Bain variants. For example, let us consider the sheared variants $B_1S_{12}(\la)$ and $B_1S_{13}(\la)$, which are macroscopically equivalent to twins between $B_1$ and $B_2$, and to twins between $B_1$ and $B_3$, respectively.\footnote{We remark that two simple shears, $B_1S_{12}(\la_1)$ and $B_1S_{13}(\la_2)$, can only be compatible if $\la_1=\la_2$.} Unlike the case of single shears (simple twins), the twins are not compound anymore and we need to distinguish between the first \eqref{EqM1} and second \eqref{EqM2} solution of Mallard's law.

Taking for instance the first solution of Mallard's law we obtain a solution triple $(R^\one,\bb^\one,\mm^\one)$ such that 
\begin{equation}\label{EqDoubleTwin}
 R^\one B_1S_{12}(\la)-B_1S_{13}(\la)=\bb^\one \tp\mm^\one.
\end{equation}
In particular, the two shears $R^\one B_1S_{12}(\la)$ and $B_1S_{13}(\la)$ can form an internal twin giving rise to a macroscopic strain $F^\one$ of the form $F^\one=(1-\mu^\one) B_1S_{13}(\la)+\mu^\one R^\one B_1S_{12}(\la)$, where $\mu^\one$ is the volume fraction of $R^\one B_1S_{12}(\la)$ in the twinning system. As before, by using \eqref{EqDoubleTwin}, $F^\one$ can be expressed as
\begin{alignat}{3}\label{EqDouble}
 F^\one= B_1S_{13}(\la)+\mu^\one \bb^\one \tp\mm^\one = B_1S_{13}(\la)(\id+\mu^\one S_{13}^{-1}(\la)B_1^{-1}\bb^\one \tp \mm^\one)=:B_1S_{13}(\la)T^\one(\mu^\one),
\end{alignat}
where $S_{13}(\la)$ is as above and $T^\one(\mu^\one)=\id+\mu^\one S_{13}^{-1}(\la)B_1^{-1}\bb^\one \tp \mm^\one$ is a shear with shearing magnitude $g^\one=|\mu^\one S_{13}^{-1}(\la)B_1^{-1}\bb^\one|$, shearing direction $\dd^\one=\mu^\one S_{13}^{-1}(\la)B_1^{-1}\bb^\one/g^\one$ and shearing normal $\mm^\one$. An analogous formula holds for the second solution of Mallard's law and for other pairs of sheared Bain variants.

We note that a deformation like $F^\one$ - see \eqref{EqDouble} - may equivalently correspond to a twin between the sheared variants $R^\one B_1S_{12}(\la)$ and $B_1S_{13}(\la)$ which, themselves, can correspond to two regions with different active slip systems in the parent, i.e. $S_{12}(\la)$ and $S_{13}(\la)$. Similarly, $F$ may also be interpreted as the macroscopic deformation corresponding to a single variant plate and two active slip systems, namely $S_{13}(\la)$ and $T^\one(\mu^\one)$. In any of the above interpretations, any interface involved would be \emph{fully coherent} due to the rank-one connections inherent in the construction of the deformation. In particular, any interface would need to be glissile \cite{Bhadeshia}.

We also remark that there are no explicit scales involved in our calculations and no prediction regarding scales can be made. Nevertheless, the model naturally contains two scales: a fast scale where the first-order twinning occurs, and a slow scale where the twin between twins occurs. Analogous two-scale phenomena apply to the other interpretations of the theory when twinning is not assumed to be the internal mechanism achieving the produced shears.

Similarly to the original (single shear) PTMC, the total shape strain $F^\one$ needs to satisfy the IPS condition, i.e. $\hat R^\one F^\one-\id=\cc^\one \tp \pp^\one$. As is well known, see e.g. \cite{Bha,BallFine,KhachBook} as well as in the Appendix, the IPS condition is equivalent to requiring that the middle eigenvalue of $(F^\one)^TF^\one$ is equal to $1$. Solving the 
IPS condition for $F^\one$ as in \eqref{EqDouble} gives a dependency $\mu^\one(\la)$. In particular, by the IPS condition, $F^\one$ in \eqref{EqDouble} only depends on a single parameter $\la$ and we can find rotations $\hat R^\one(\la)$ such that 
\begin{equation}\label{IPS}
\hat R^\one(\la) F^\one(\la)=\hat R^\one(\la) B_1S_{13}(\la)T^\one(\mu^\one(\la))=\id + \cc^\one(\la) \tp \pp^\one(\la).
\end{equation}

Using this approach, one can construct a double shear theory which is macroscopically equivalent to (double) twinning. We will show that this choice of shearing systems naturally results in a theory that predicts the formation of certain \nono{h}{h}{k} habit planes in low carbon steels, such as \nono{5}{5}{7} or \nono{2}{2}{3}.

In \eqref{IPS} we have computed a $\la$ dependent family of double shears of $B_1$ that uses the Type $\one$ solution for the second shearing system. Similarly, we can construct double shears of any Bain variant, using either the Type $\one$ or Type $\two$ solution for the second shearing system. Furthermore, we note that by Proposition \ref{PropBJ} in the Appendix, for each $F^\one(\la)$ and each $F^\two(\la)$ there exist exactly two rotations and two shearing systems that satisfy the IPS condition. To avoid overcomplicating notation, we will not explicitly distinguish between these two solutions.

\begin{figure}[h]
 \centering
\begin{overpic}[width=\columnwidth]{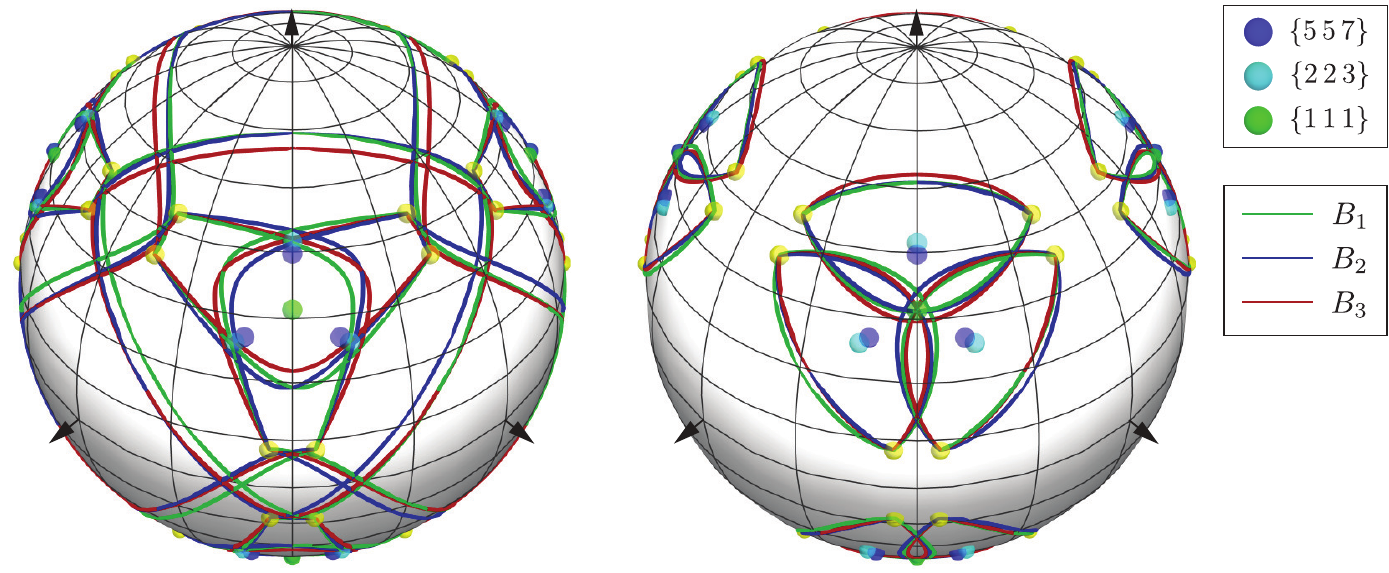}
 \put(-1.2,35){$F^{\one}(\la)$}
 \put(43,35){$F^{\two}(\la)$}
\end{overpic}
  \caption{{Possible habit plane normals for $F^{\one}(\la)$ and $F^{\two}({\la})$ for $\Delta V = 0$ plotted on the unit sphere. Differently colored lines (green, blue, red) correspond to the three different Bain variants being sheared and each point on a colored line corresponds exactly to one possible habit plane normal. Yellow dots correspond to habit planes arising from single shear theories. For reference, points corresponding to \nono{5}{5}{7},\nono{2}{2}{3} and \nono{1}{1}{1} habit planes are also shown.}}   
  \label{FigNorm}
\end{figure}

{A plot of the resulting habit plane normals for all Bain variants, both types of solutions and both solutions for the IPS condition is shown in Figure \ref{FigNorm}. The unit sphere represents the space of all possible (normalized) habit plane normals. Due to the intrinsic $\PP$ symmetry of the problem, $x,y,z$ directions are only indicated by arrows but not explicitly labeled as any permutation of the axes would leave the plot unchanged.} In this plot the fcc and bcc lattice parameters have been chosen such that no volume change occurs, i.e. $\Delta V = 0$. For volume changes of up to $\pm 1\%$, this plot remains qualitatively similar. We refer the reader to Section \ref{Secresults} for examples with different lattice parameters.

In this figure, the yellow points represent habit planes that arose from single shear theories. For highly tetragonal steels, these are precisely the habit planes in the family \nono{3}{10}{15} in the original Wechsler, Liebermann and Read (single shear) PTMC. Also, taking the first solution $F^\one$, one can see that there is a high density of intersections of all the differently colored lines, i.e. the differently sheared Bain variants, very close to \nono{5}{5}{7} or \nono{2}{2}{3}. For the second solution $F^\two$, the density of intersections is highest very close to \nono{1}{1}{1}.

\subsection{Selection mechanism for double shear theories}\label{SecSel}
So far, we have established double shear theories that only depend on a single parameter $\la$. In this section, out of this family of possible double shear theories, we identify parameters $\la$ that give rise to double shear theories that satisfy a criterion of small shape strain magnitude and maximal compatibility. We will see that double shear theories resulting in near \nono{5}{5}{7} habit planes satisfy both criteria and are thus preferable.

\subsubsection*{Criterion of small shape strain magnitude:}

\begin{figure}[h]
  \centering
  \begin{overpic}[width=.95\columnwidth]{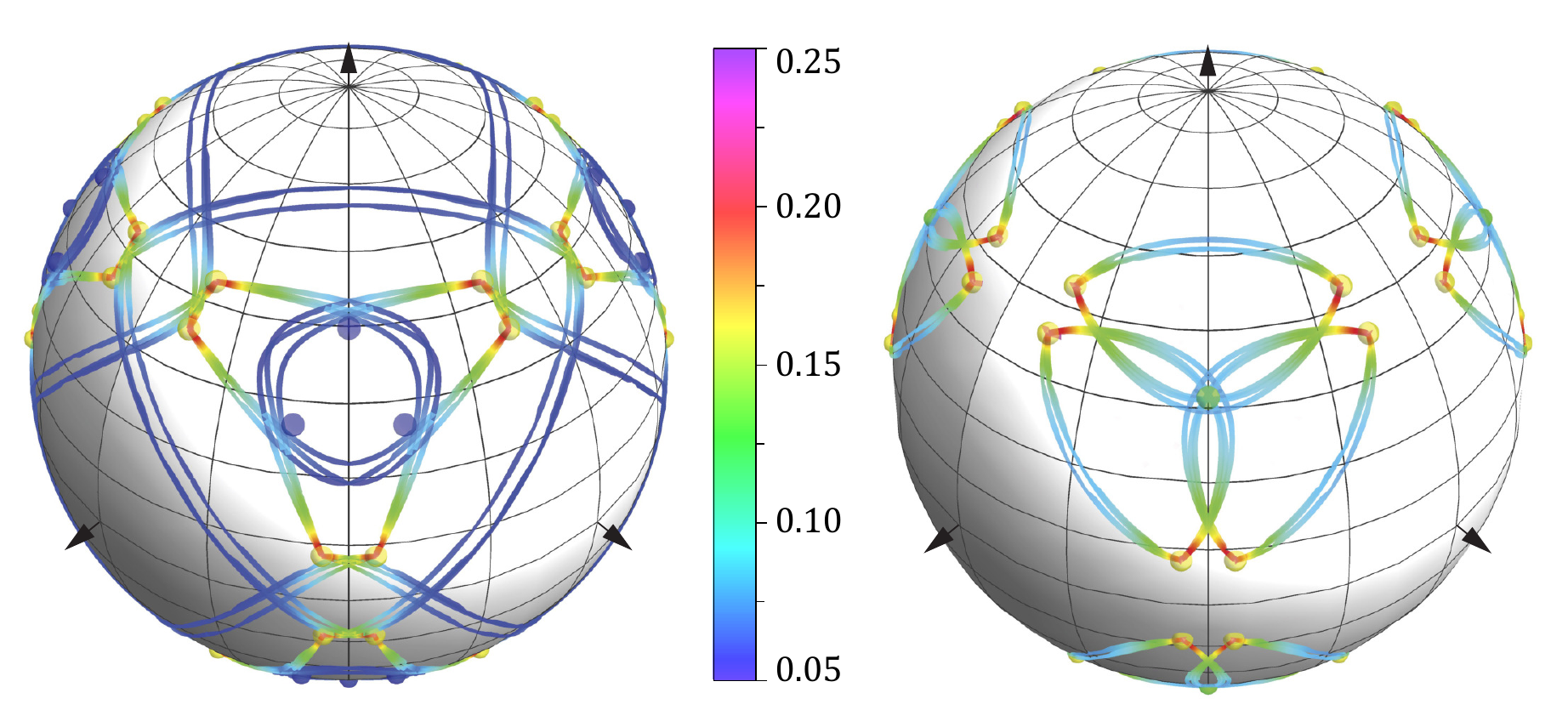}
 \put(2,41){$F^{\one}(\la)$}
 \put(57,41){$F^{\two}(\la)$}
\end{overpic}
  \caption{Plot of possible habit plane normals for $F^{\one}(\la)$ and $F^{\two}({\la})$ for $\Delta V = 0$ on the unit sphere. The color gradient indicates the magnitude of the shape strain of $F^{\one}(\la)$ and $F^{\two}({\la})$, respectively.}  
  \label{FigStrains}
\end{figure}

Following \cite{Kelly}, see also \cite{MoritoShapeStrain}, we seek double shear theories that result in a low shape strain magnitude $|\cc|$ for a shape strain of the form $RF=\id + \cc\tp\pp$ as in \eqref{IPS}. Figure \ref{FigStrains} repeats the plot of habit planes from Figure \ref{FigNorm} but this time assigns a color gradient depending on the shape strain magnitude of the corresponding $F$.

Firstly, it can be seen from Figure \ref{FigStrains} that single shear theories (yellow points) give rise to the highest shape strain magnitudes as expected. We note that single shear theories correspond to the following cases (cf. \eqref{EqDouble}):
\begin{itemize}
 \item if $\mu=0$ then $F^\one= B_1S_{13}(\la)$ and thus a single shear theory for $B_1$,
 \item if $\mu=1$ then $F^\one= R^\one B_1S_{12}(\la)$ by \eqref{EqDoubleTwin} and thus a single shear theory for $B_1$, and
 \item if $\la=1$ then, using that $B_1S_{13}(1)=RB_3$ for some rotation $R$, we obtain $F^\one=RB_3T^\one(\mu)$ and thus a single shear theory for $B_3$.
\end{itemize}
The same classification applies to the second solution $F^\two$. Starting from any of the single shear theories there is a one-parameter family of double shear theories that connects it to another single shear theory. For each double shear theory, we can compute the corresponding habit plane. The resulting arcs of habit planes connecting two single theories are shown in Figure \ref{FigStrains}.\footnote{The habit planes corresponding to single shear theories are indicated by yellow dots in Figure \ref{FigStrains}.} The coloring of the normals indicates the shape strain magnitude of the underlying double shear theory. It can be seen that this magnitude gets smaller the further away a double shear theory is from a single shear theory. 

It can further be seen from Figure \ref{FigStrains} that the smallest shape strain magnitudes are obtained from double shear theories of Type I and that the smallest shape strain magnitudes for solutions of Type II are achieved near the point of highest density of intersections. Even though the shape strain magnitude of double shear theories resulting in near $\{5\,5\,7\}$ habit planes is not minimal, it is low and in particular lower than that of any double shear theory of Type II.

The obtained shape strain values lie within the range of 0.07 to 0.08 which is relatively low for a single-variant lath, compared to data reporting values in the range 0.1 - 0.3, see \cite{Bryans} for FeNi or \cite{Wakasa81} for FeNiMn.

\subsubsection*{Criterion of maximal compatibility:}
Apart from low shape strain magnitudes, points near \nono{5}{5}{7} and \nono{1}{1}{1} also have a high density of intersections and hence satisfy a strong criterion of compatibility. On a very basic level, having a high density of intersections simply implies that there are more double shear theories that give rise to habit planes close to \nono{5}{5}{7} and \nono{1}{1}{1}. 

Furthermore, suppose that two regions within the same prior austenite grain have been deformed according to two different double shear theories which share the same habit plane $\pp$. Denoting the respective shape strains in the two regions by $\id+\cc_1\tp \pp $ and $\id+\cc_2\tp \pp$, we observe that 
\begin{equation}\nonumber
 (\id+\cc_1\tp \pp) - (\id+\cc_2\tp \pp) = (\cc_1 - \cc_2)\tp \pp.
\end{equation}
Thus, any two such regions can share a fully coherent interface of normal $\pp$.
In particular, the most likely double shear theories with this property are the ones with habit planes near \nono{5}{5}{7} and \nono{1}{1}{1}. This compatibility property may play a crucial role when thinking of the dynamic process of nucleation. As austenite is quenched, the martensite phase nucleates at various sites as an IPS with all three Bain variants being equally likely to occur. As explained above, if two growing nuclei happen to share the same habit plane $\pp$, they are able to meet along a fully coherent interface. We note that in Figure~\ref{FigNorm}, the different colors correspond to double shears of different Bain variants and thus if three differently colored lines intersect at one point it implies that there are double shear theories of all three Bain variants that share this habit plane. Remarkably, both points near \nono{5}{5}{7} and \nono{1}{1}{1} share this property. 

\section{Results}\label{Secresults}

We apply our model to explore the formation of \nono{h}{h}{k} habit planes with the volume change parameters $\Delta V$ derived in Section \ref{SecThermal} and also investigate the resulting orientation relationships (ORs). Here, we present the results for the double shear theories with the lowest and highest volume changes which, as we shall see, result in habit planes near \nono{5}{5}{7} and ORs near the Kurdjumov-Sachs model. Results for different lattice parameters can be obtained with the help of the MATLAB App \emph{``Lath Martensite''} available at\texttt{\href{https://github.com/AntonMu/LathApp}{github.com/AntonMu/LathApp}}.

In order to determine double shear theories satisfying the criterion of maximal compatibility and low shape strain magnitude, we first calculate all possible families of habit planes. We then search for all points of intersections between these families that have low shape strain magnitude. In all cases, we find that the two criteria are satisfied near points in the family \nono{h}{h}{k}. A reason for finding a high density of intersections near such habit planes may be that any double shear theory resulting in a \nono{h}{h}{k} normal trivially intersects with at least one of its crystallographically equivalent families. This is because if the family of shape strains $F(\la)$ results in a habit plane $\pp(\la^*)=\no{h}{h}{k}$ for some $\la^*$, then there exists an element $P$ in the cubic point group, i.e. an orthogonal transformation mapping the cube to itself, such that $P \no{h}{h}{k}=\no{h}{h}{k}$. In particular, the crystallographically equivalent families of double shear theories given by $F(\la)$ and $PF(\la)P^T$ intersect at \no{h}{h}{k}. 

We recall that, for $RF$ as given in \eqref{IPS}, the $P$-crystallographically equivalent double shear system is given by
\begin{equation}\label{EqCrysEq}
 PRFP^T = (PR P^T)(PB_1P^T)(PS_{13}P^T)(PTP^T)=\id + (P\cc) \tp (P\pp),
\end{equation}
which is a double shear theory for $PB_1P^T$ with habit plane $P\pp$ and a crystallographically equivalent orientation relationship. 

\begin{table}[h!]
\begin{center}
  \begin{tabular}{r|cccc}
$\la^*$ & $\mu^\one(\la^*)$ & $S(\la^*)$ & $T^\one(\mu^\one(\la^*))$ & $R^\one (\la^*)F^\one (\la^*) $ \\ 
\hline \\[-.85em]
$.5772$ 
& $.5717$
& $.3848 \vec{\hphantom{-}.707}{-.707}{0}\noc{\hphantom-.707}{\hphantom-.707}{0}$ 
& $.2332 \vec{\hphantom{-}.226}{\hphantom-.689}{\hphantom-.689}\noc{\hphantom-0}{\hphantom-.707}{-.707}$ 
& $.0728 \vec{-.836}{\hphantom{-}.548}{-.012}\noc{\hphantom-.494}{\hphantom-.715}{\hphantom-.494}$ \\[1.5em]
 $.6609$
& $.6263$
& $.4406 \vec{\hphantom{-}.707}{-.707}{0}\noc{\hphantom-.707}{\hphantom-.707}{0}$
& $.2885 \vec{\hphantom{-}.177}{\hphantom-.696}{\hphantom-.696}\noc{\hphantom-0}{\hphantom-.707}{-.707}$
& $.0723 \vec{-.048}{-.533}{-.847}\noc{\hphantom-.474}{-.742}{\hphantom-.474}$ \\[1.5em]
$.7602$
 & $.5496$
 & $.5068 \vec{\hphantom{-}.707}{-.707}{0}\noc{\hphantom-.707}{\hphantom-.707}{0}$ 
& $.2869 \vec{\hphantom{-}.122}{\hphantom-.702}{\hphantom-.702}\noc{\hphantom-0}{\hphantom-.707}{-.707}$ 
& $.0777 \vec{\hphantom-.585}{\hphantom{-}.124}{-.802}\noc{\hphantom-.726}{-.486}{\hphantom-.486}$ 
\end{tabular}
\caption{Elements of the double shear system \eqref{IPS} leading to habit planes near \nono{5}{5}{7} for $\Delta V_{\mathrm{Fe-0.6C}}(1040K) = -0.2\%$. The remaining \nono{5}{5}{7} habit planes can be obtained from the crystallographically equivalent systems. Shearing directions and normals are of unit length. We use the shorthand notation $g\, \dd\, \kk$ for $\id + g\, \dd \tp \kk$. {All vectors are expressed in $\gamma$ coordinates.}} \label{TableShearNeg}
\end{center}
\end{table}

\subsection{Resulting Habit Planes}
\label{sec:3.1}

In Table \ref{TableShearNeg}, we use the lowest volume change of $\Delta V_{\mathrm{Fe-0.6C}}(1040K)=-0.2\%$ following the calculation in \cite{Onink}. Using the Type I shearing system in \eqref{EqDouble} for e.g. $\la^*=0.5772$, the IPS condition is satisfied for $\mu^\one(\la^*)=0.5717$ resulting in the double shear theory $R^\one(\la^*) F^\one(\la^*)= R^\one(\la^*) B_1S(\la^*)T^\one(\mu^\one(\la^*))=\id + \cc^\one(\la^*) \tp \pp^\one(\la^*)$. Similarly, in Table \ref{TableShearPos}, we use $\Delta V_{\mathrm{Fe-0.252C}}(1042K)=0.865\%$ (cf. \cite{FewSteels}[Sample S556]) which is the average of the possible volume changes in the transformation temperature range for Fe-0.252C. 

\begin{table}[h!]
\begin{center}
  \begin{tabular}{r|cccc}
$\la^*$ & $\mu^\one(\la^*)$ & $S(\la^*)$ & $T(\mu^\one(\la^*))$ & $R^\one (\la^*)F^\one (\la^*) $ \\ 
\hline \\[-.85em]
$.5717$ 
& $.6157$
& $.3811 \vec{\hphantom{-}.707}{-.707}{0}\noc{\hphantom-.707}{\hphantom-.707}{0}$ 
& $.2490 \vec{\hphantom{-}.229}{\hphantom-.688}{\hphantom-.688}\noc{\hphantom-0}{\hphantom-.707}{-.707}$ 
& $.0840 \vec{-.745}{\hphantom{-}.656}{-.120}\noc{\hphantom-.459}{\hphantom-.761}{\hphantom-.459}$ \\[1.5em]
 $.6402$
& $.6572$
& $.4268 \vec{\hphantom{-}.707}{-.707}{0}\noc{\hphantom-.707}{\hphantom-.707}{0}$
& $.2942 \vec{\hphantom{-}.189}{\hphantom-.694}{\hphantom-.694}\noc{\hphantom-0}{\hphantom-.707}{-.707}$
& $.0839 \vec{-.146}{-.641}{-.753}\noc{\hphantom-.442}{-.781}{\hphantom-.442}$ \\[1.5em]
$.7876$
 & $.5392$
 & $.5251 \vec{\hphantom{-}.707}{-.707}{0}\noc{\hphantom-.707}{\hphantom-.707}{0}$ 
& $.2906 \vec{\hphantom{-}.107}{\hphantom-.703}{\hphantom-.703}\noc{\hphantom-0}{\hphantom-.707}{-.707}$ 
& $.0902 \vec{\hphantom-.670}{\hphantom{-}.203}{-.714}\noc{\hphantom-.766}{-.455}{\hphantom-.455}$ 
\end{tabular}
\caption{Elements of the double shear system \eqref{IPS} leading to habit planes near \nono{5}{5}{7} for $\Delta V_{\mathrm{Fe-0.252C}}(1042K) = 0.865\%$. The remaining \nono{5}{5}{7} habit planes can be obtained from the crystallographically equivalent systems. Shearing directions and normals are of unit length. We use the shorthand notation $g\, \dd\, \kk$ for $\id + g\, \dd \tp \kk$. {All vectors are expressed in $\gamma$ coordinates.}} \label{TableShearPos}
\end{center}
\end{table}

As pointed out in Section \ref{SecSel}, the shape strains for Type II solutions, resulting in \nono{1}{1}{1} habit planes for $0\%$ volume change, have higher shape strain magnitude. For volume changes above $\approx0.6\%$ the resulting habit planes fail to intersect and thus do not satisfy the criterion of maximal compatibility. For, potentially hypothetical, negative volume changes below $\approx-0.6\%$, also the shape strains for Type II solutions result in habit planes near \nono{5}{5}{7}. These and similar observations can be obtained easily with the MATLAB App \href{https://github.com/AntonMu/LathApp}{\emph{``Lath Martensite''}}.

\subsection{Resulting Orientation Relationships}

In this section we establish the orientation relationships (ORs) for the double shear theories obtained in Tables \ref{TableShearNeg} and \ref{TableShearPos}. In particular, we restrict our investigation to Type I solutions and, for the ease of the reader, we henceforth omit the superscript $I$ in our notation. For these solutions, we derive the 
resulting ORs and compare them to typical ORs stated in the literature, that is to the Nishiyama--Wassermann (\NW), Kurdjumov--Sachs (\KS), Pitsch (\PT) and Greninger--Troiano (\GT) orientation relationships.

We derive the OR based on the overall rotation $R$ that is needed to make the (double) sheared Bain strain $B_1$ in \eqref{IPS} an invariant plane strain $RF = RB_1ST = 1 + \cc\otimes\pp$. Following \cite{KSPaper}, the corresponding OR matrix $O_{\mathrm{tot}}$ transforming the fcc to the bcc basis, is then given by
\begin{equation}\label{EqOR}
    O_{\mathrm{tot}} = R[-45^\circ,\ee_1]R^T.
\end{equation}
 Here, $R[\theta,\ee_1]$ denotes the rotation about the unit vector $\ee_1=[1,0,0]$ by an angle $\theta$. We recall that the roation $R[\theta,\ee_1]$ stems from the change of coordinates when changing from an fcc to a bcc basis with the pure Bain strain $B_1$. Since the OR matrix is a change of basis matrix from fcc to bcc, it maps fcc normals \no{n_1}{n_2}{n_3} to bcc normals  \noa{\hat n_1}{\hat n_2}{\hat n_3} and fcc directions \ve{v_1}{v_2}{v_3} to bcc directions \vea{\hat v_1}{\hat v_2}{\hat v_3}. For example, the orientation relationship \KS22 is characterized by $O_{\KS22}$ such that
 \begin{equation}\nonumber
O_{\KS22}\no{\bar 1}{\bar 1}{1}=\noa{\bar 1}{0}{1} \mbox{ and } O_{\KS22}\ve{1}{\bar 1}{0}=\vea{1}{\bar 1}{1},
 \end{equation}
where $O_{\KS22}=R[-45^\circ,\ee_1]R[-9.74^\circ,\ve{0 }{1}{1}]R[-5.26^\circ, \ve{\bar 1}{\bar 1}{1}]$.\footnote{See \cite{KSPaper} for the exact numerical values of the angles.} Equivalently, this relationship can be expressed in terms of the parallelisms $\no{\bar 1}{\bar 1}{1}\,\|\,\noa{\bar 1}{0}{1}$ and $\ve{1}{\bar 1}{0}\,\|\,\vea{1}{\bar 1}{1}$. To determine which common OR is closest to $O_{\mathrm{tot}}$ we use two different approaches: 
\begin{enumerate}
    \item \textbf{Minimising the Relative Rotation}\\
    Given $O_{\mathrm{tot}}$ as in \eqref{EqOR}, we find an OR $O_*$ such that the relative rotation $R_{\mathrm{rel}}$, where  $O_{\mathrm{tot}}=R_{\mathrm{rel}} O_*$, is minimal. Here, $*$ denotes any of the possible models \NW, \KS, \PT\ or \GT, which also undergo a transformation according to $B_1$, i.e. are of the form $O_*=R[-45^\circ,\ee_1] R_*$. 
    \item \textbf{Minimising Angular Deviations of Planes and Directions}\\
    Given $O_{\mathrm{tot}}$ as in \eqref{EqOR}, we compare known parallisms between fcc and bcc planes and directions found in common ORs and calculate the angular deviations from them. For instance, when considering $\KS22$, we compute
\begin{equation}\nonumber
  \angle{\left(O_{\mathrm{tot}}\no{\bar 1}{\bar 1}{1},\noa{\bar 1}{0}{1}\right)} \aand \angle{\left(O_{\mathrm{tot}}\ve{1}{\bar 1}{0},\vea{1}{\bar 1}{1}\right)}.
\end{equation}
We will use the short-hand notation $\no{\bar 1}{\bar 1}{1}:\noa{\bar 1}{0}{1}=\angle{\left(O_{\mathrm{tot}}\no{\bar 1}{\bar 1}{1},\noa{\bar 1}{0}{1}\right)}$ to express this relationship. As before we restrict the candidate ORs to those that undergo a transformation according to $B_1$. The minimisation is then performed over the sum of the deviation angles between normals and directions. 
\end{enumerate}

By performing both minimisations for the previous examples of Fe-$0.6$C (cf. Table \ref{TableShearNeg}) and Fe-$0.252$C (cf. Table \ref{TableShearPos}) we find that $O_{\mathrm{tot}}$ is closest to $\KS22$ according to both methods for all solutions. The deviations from $\KS22$ are summarised in Tables \ref{TableShearNegOR} and \ref{TableShearPosOR} below. {We note that the resulting ORs are different for each of the computed double shear theories but are nonetheless all closest to $\KS22$. }

\begin{table}
\begin{center}
\begin{tabular}{r|ccc}
$\la^*$ & $\no{\bar 1}{\bar 1}{1}:\noa{\bar 1}{0}{1}$ & $\ve{1}{\bar 1}{0}:\vea{1}{\bar 1}{1}$ &$R^\one (\la^*)R_{\KS22}^T$\\ 
\hline \\[-.85em]
$.5772$ 
& $\approx0.94^\circ$
& $\approx6.52^\circ$ 
& $R[6.54^\circ,  \vecomma{-.502}{-.524}{.688}]$\\
 $.6609$
& $\approx1.52^\circ$
& $\approx4.75^\circ$
& $R[4.79^\circ,  \vecomma{-.517}{-.340}{.786}]$\\
 $.7602$
& $\approx4.45^\circ$
& $\approx7.47^\circ$
& $R[7.74^\circ,  \vecomma{-.446}{-.079}{.892}]$
\end{tabular}

\caption{Angular deviations and relative rotations between \KS22 and the orientation relationships resulting from the double shear system from Table \ref{TableShearNeg} for $\Delta V_{\mathrm{Fe-0.6C}}(1040K) = -0.2\%$. $\no{\bar 1}{\bar 1}{1}:\noa{\bar 1}{0}{1}\approx0.94^\circ$ indicates that the normals deviate approximately $0.94^\circ$ from being parallel or equivalently that $\angle{\left(O_{\mathrm{tot}}\no{\bar 1}{\bar 1}{1},\noa{\bar 1}{0}{1}\right)}\approx0.94^\circ$. $R[6.54^\circ,  \vecomma{-.502}{-.524}{.688}]$ denotes a rotation of $6.54^\circ$ about $\vecomma{-.502}{-.524}{.688}$ and indicates the relative rotation between $R^\one (\la^*)$ and $R_{\KS22}$.} \label{TableShearNegOR}
\end{center}
\end{table}

\begin{table}
\begin{center}
  \begin{tabular}{r|ccc}
$\la^*$ & $\no{\bar 1}{\bar 1}{1}:\noa{\bar 1}{0}{1}$ & $\ve{1}{\bar 1}{0}:\vea{1}{\bar 1}{1}$ &$R^\one (\la^*)R_{\KS22}^T$ \\ 
\hline \\[-.85em]
$.5717$ 
& $\approx1.12^\circ$
& $\approx7.00^\circ$ 
& $R[7.03^\circ,  \vecomma{-.579}{-.454}{.677}$] \\
 $.6402$
& $\approx0.97^\circ$
& $\approx4.22^\circ$
& $R[4.25^\circ,  \vecomma{-.559}{-.402}{.725}]$\\
 $.7876$
& $\approx5.2^\circ$
& $\approx8.23^\circ$
& $R[8.56^\circ,  \vecomma{-.429}{-.044}{.902}]$
\end{tabular}

\caption{
Angular deviations and relative rotations between \KS22 and the orientation relationships resulting from the double shear system from Table \ref{TableShearPos} for $\Delta V_{\mathrm{Fe-0.252C}}(1042K) = 0.865\%$. $\no{\bar 1}{\bar 1}{1}:\noa{\bar 1}{0}{1}\approx1.12^\circ$ indicates that the normals deviate approximately $1.12^\circ$ from being parallel or equivalently that $\angle{\left(O_{\mathrm{tot}}\no{\bar 1}{\bar 1}{1},\noa{\bar 1}{0}{1}\right)}\approx1.12^\circ$. $R[7.03^\circ,\vecomma{-.579}{-.454}{.677}]$ denotes a rotation of $7.03^\circ$ about $\vecomma{-.579}{-.454}{.677}$ and indicates the relative rotation between $R^\one (\la^*)$ and $R_{\KS22}$.}
 \label{TableShearPosOR}
\end{center}
\end{table}
Recalling that by \eqref{EqCrysEq} all $P$-crystallographically equivalent double shear systems are given by $PRFP^T$, we readily conclude that the $P$-crystallographically equivalent OR $PO_{\mathrm{tot}}P^T$ is closest to the orientation relationship $P O_{\KS22}P^T$ which corresponds to a different \KS variant.

Finally, we remark that the algebraic complexity of the problem prohibits us from deriving closed form expressions that show that 
$O_{\mathrm{tot}}$ is always closest to a \KS variant. However, numerical evidence obtained through the MATLAB App \href{https://github.com/AntonMu/LathApp}{\emph{``Lath Martensite''}} provides strong evidence that this claim holds true for all physical ranges of volume changes that can be observed in fcc to bcc phase transformations for Type I, and also Type II, solutions. Near \KS orientation relationships are commonly reported in the literature \cite{Morito,MoritoShapeStrain}. In \cite{MoritoShapeStrain} the authors analyse EBSD patterns in SEM as well as the more accurate Kikuchi diffraction patterns in TEM for several FeC alloys with varying carbon content, concluding that the dominant orientation relationship is near \KS. We note that intermediate orientations between \KS and \NW (\GT) are also reported, e.g. in the study of the orientation between laths and narrow films of retained austenite in low carbon steels \cite{Kelly90}.

\section{Conclusion \& Outlook}

In analogy to and motivated by the groundbreaking work of Wechsler, Liebermann \& Read and Bowles \& Mackenzie on the original PTMC, we have built a double shear theory by choosing the shearing systems as those arising from (second order) twinning.

Our developed double shear theory agrees well with the experimental observation of near \nono{5}{5}{7} habit planes in low-carbon steels and results in orientation relationships close to Kurdjumov-Sachs for a single lath. 

Unlike some existing double shear theories based on PTMC, no parameter fitting and/or parameter estimation was necessary to reach our results. By choosing shearing systems that are macroscopically compatible with twinning and requiring that the overall shape strain is an invariant plane strain, our model \emph{predicts} near \nono{5}{5}{7} habit planes solely based on the additional assumptions of small shape strain magnitude and a condition of maximal compatibility. We remark that, as other double shear theories, our theory is macroscopic and does not imply, or necessitate, internal twinning. In fact, it is compatible with different interpretations of the internal structure: double twinning, double slip, or twinning between regions with different slip systems.

Furthermore, as e.g. in \cite{Maresca}, our theory reveals a very sensitive dependence of the possible lath habit planes on the volume change during transformation from fcc austenite to bcc martensite. It would be interesting to put this theoretical dependency to experimental scrutiny. 

\section*{Acknowledgements}
The research of A. M. leading to these results has received funding from the German Academic Exchange Service and the FEAP project at the University of California, Berkeley. The authors wish to thank S. Govindjee, J.W. Morris, Jr. and A. Minor for helpful discussion during the development of this paper. 
\section*{Appendix}
The following proposition gives necessary and sufficient conditions for a $3\times 3$ matrix $F$ to be an invariant plane strain, up to a rotation $R$.

\begin{Proposition}{\cite{BallFine}}\label{PropBJ}
 Let $F$ be a  $3\times 3$ matrix such that $F^TF\neq \id$, i.e. $F$ is not a pure rotation, and let $\lambda_1 \leq \lambda_2 \leq \lambda_3$ be the ordered eigenvalues of $F^TF$. Then, there exist a rotation $R$ and vectors $\bb, \mm$ such that
 \begin{equation*} 
  RF=\id+\bb \tp \mm  
 \end{equation*}
if and only if $\lambda_1 \geq 0$ and $\lambda_2=1$. If these conditions are satisfied, there are at most two solutions given by
 \begin{align*}
  &\bb=\frac{\rho}{\sqrt{\lambda_1^{-1}-\lambda_3^{-1}}}\left( \sqrt{\lambda_1^{-1}-1}{\bf v}_1+\kappa \sqrt{1-\lambda_3^{-1}} {\bf v}_3 \right) ,  \\
  &\mm=\rho^{-1} \left( \frac{\sqrt{\lambda_3}-\sqrt{\lambda_1}}{\sqrt{\lambda_3-\lambda_1}}\right) (- \sqrt{1-\lambda_1}{\bf v}_1+\kappa \sqrt{\lambda_3-1}{\bf v}_3),  
 \end{align*}
 where $\rho \neq 0$ is chosen to make $\mm$ of unit length, $\kappa \in \lbrace -1,1\rbrace$ and ${\bf v}_1,{\bf v}_3$ are the (normalized) eigenvectors of $F^TF$ corresponding to $\lambda_1$ and $\lambda_3$.
\end{Proposition}

\bibliography{Bib}
\bibliographystyle{alpha}

\end{document}